# ABOUT SELF ENERGY IN 3-DIMENSIONAL ELECTRODYNAMICS.

# M. SH. PEVZNER.

Department of Physics, National Mining University of Ukraine,

19, Karl Marx Avenu, Dniepropetrovsk, 49600, Ukraine

E-mail: PevznerM@nmuu.dp.ua, mark@omp.dp.ua

Abstract

The fermion electromagnetic mass has been evaluated in 3-dimensional electrodynamics in the first nonvanishing order of $N^{-1}$ approximation ($N$ is the number of the fermions in the model). It is established, that, if one proceeds from the self energy integral in QED$_3$, in the approximation under consideration for the mentioned above mass one can obtain the quantum result for the fermion electromagnetic mass as well the classical one. The electromagnetic mass of the particle has the quantum origin, if its Compton wavelength satisfies the condition $\lambda_c \gg a$ ($a$ is the scale parameters of the model which are the "elementary length" $r_0$ and $8/\alpha$, $\alpha = e_0^2 N$, $e_0^2$ is the dimensional coupling constant in QED$_3$); the particle electromagnetic mass has the classical origin, if $\lambda_c \ll a$. The obtained results are discussed.

---------------------------------

1. The problem of a particle self energy is one of unsolved problems both in classical and in quantum physics hitherto. In particular it was considered for a long time that in the electrodynamics both the classical and quantum self energy have a different origin [1]. It was also estimated that the classical approach to this problem is meaningless since the result obtained on the base of the classical approach is not a limiting case of the quantum one. As the electrical radius of the electron satisfies the unequality $r_{ел} \ll \lambda_c$ ($\lambda_c$ is the Compton wavelengh of the electron) the quantum effects connected with the particle interaction with its own field are felt at much larger distances than classical ones what at the first sight makes unexpedient the classical approach to the problem mentioned above.

In works [2,3] the classical approach to the problem of the fermion mass origin has been partially reabilitated. In particular it was shown that in QED$_4$ some approximations allow to obtain from the fermion self energy quantum integral both the classical and quantum expressions for the electromagnetic fermion mass. Hereat the unequality $\lambda_c \gg r_0$ corresponds to the quantum asymptotic of the denoted integral ($\delta m \sim \ln 1/r_0$), then to the classical asymptotic ($\delta m \sim 1/r_0$) corresponds the unequality $\lambda_c \ll r_0$ where $r_0$ is the "elementary length".

2. This work is devoted to the evaluation of the fermion mass in three-dimensional electrodynamics (QED$_3$) and to investigation of connection between the fermion mass obtained by the classical and quantum approaches. There are some particularities here in comparison with the four dimensions case. Firstly there is the dimensional coupling constant $e_0^2$ that introduces the additional scale of distances. Secondly if one evaluates fermion self energy by the classical method the corresponding integral upon momenta diverges both at the lower and at the upper bounds. The divergence at the upper bound may be removed by the ordinary introducing of cutoff momentum $k_{max} = 1/r_0$; as to the divergence at the lower bound in the case under consideration if one does not introduce the photon mass it is necessary to use infrared cutoff to which it is difficult to ascribe any satisfactory physical meaning. In other words the completely classical solving of the inconviencies mentioned above does not exist. One of the ways to overcoming the problems encounting here is shown in [4,5]: when one evaluates the point source energy by classical manner, one has to deal with the field potential taking into account vacuum polarization. The limiting by the first nonvanishing term of the $N^1$ - expansion of the polarization operator ($N$ is a fermion number in the model) allows to avoid the divergence in the classical expression for the self energy at the small momenta if the dynamical fermions masses are equal to zero. In this case the screening of the point charge field takes place at the distances $r \gg 8/\alpha$ (where $\alpha = e_0^2 N$).

First the attention of the importance of the dynamical fermions masses presence for asymptotical behaviour of the point charge field potential has been turned to in the works [6,7] (this problem has been discussed in detail in [8] as well). The physical meaning of the point charge field screening if the dynamical fermions masses are equal to zero may be briefly explained in such a way. In case of the massless dynamical fermions the photon propagator taken with preserving only the first nonvanishing term of $N^1$ - expansion has no photon pole at $k^2 = 0$; at this point the photon propagator has a branch and in the spacelike momenta domain it has the pole at $k^2 = (8/\alpha)^2$. Then the single - photonic state with $k^2 = 0$ is unstable in given model that violates the confinement and leads to the screening of the point charge field in QED$_3$.

3. Evaluating the fermion dynamical mass we shall start from the self energy integral

$$\Sigma(p) = \frac{e_0^2}{(2\pi)^3} \int \gamma_\mu G(p+k) \Gamma_\nu(p+k,p) D_{\mu\nu}(k) d^3k, \qquad (1)$$

where $G$ is the full fermion propagator, $D_{\mu\nu}$ is the photon propagator and $\Gamma_\nu$ is the full vertex.

. Let us try now to find the fermion electromagnetic mass bounding ourselves for the first nonvanishing term of the $N^1$ - expansions of the quantities $G$, $D_{\mu\nu}$ та $\Gamma_\nu$. Then we have [9]

$$G(p+k) = -\frac{1}{i} \frac{i(\hat{p}+\hat{k}) - m_0}{(p+k)^2 + m_0^2}, \qquad (2)$$

$$\Gamma_\nu = \gamma_\nu, \qquad (3)$$

$$D^{(k)}_{\mu\nu} = \left(\delta_{\mu\nu} - \frac{k_\mu k_\nu}{k^2}\right) D^{(t)}(k^2) + \frac{k_\mu k_\nu}{k^2} D^{(l)}(k^2), \qquad (4)$$

$$D^{(t)}(k^2) = \frac{1}{i} \frac{1}{k^2 + (\alpha/8)k}, \qquad (4a)$$

here $m_0$ is the "bare" fermion mass, $D^l(k^2)$ is an arbitrary fuction which fixes the potential gauge.

When the relations (2) – (4) were inserted into the equality (1) we obtain

$$\Sigma(p) = -\frac{\alpha}{(2\pi)^3 N} \times$$

$$\times \int \frac{\left(2im_0 - 2\hat{k} - 2\hat{k}(pk)/k^2\right) D^{(t)}(k^2) + \left(im_0 + (\hat{k} - \hat{p}) + 2k(pk)/k^2\right) D^{(l)}(k^2)}{(p+k)^2 + m_0^2} d^3k \qquad (5)$$

(we emphasize that if $N^{-1}$ — expansion is used $\alpha$ is considered as a constant and $N^{-1}$ is considered as the smallness parameter).

For the fermion physical mass $m = m_0 + \delta m$ and when one finds the expression for the electromagnetic mass at the first nonvanishing order upon $N^{-1}$ we neglect the difference between $m$ and $m_0$ under the integral (5) and go to the coordinate system in which $\vec{p} = 0$. Then we have

$$\delta m = i\Sigma(\hat{p} = im) =$$

$$= \frac{im\alpha}{(2\pi)^3 N} \int \left[ 2\left(\frac{k_0^2}{k^2} + 1 - \frac{k_0}{m}\right) D^{(t)}(k^2) - \left(2\frac{k_0^2}{k^2} + 1 - \frac{k_0}{m}\right) D^{(l)}(k^2) \right] \frac{d^3k}{k^2 - 2mk_0}. \qquad (6)$$

The transfer to euclide variables $k_e(\vec{k}, k_0 \to ik_0')$ and the following integration upon the angles allow to give the following form to the relation (6)

$$\delta m = \frac{\alpha}{(2\pi)^2 Nm} \int_0^{k_{max}} \left[ D^{(t)}(k_e^2) \left(1 + \frac{4m^2 - k_e^2}{2mk_e} \arctg \frac{2m}{k_e}\right) + \frac{m}{k_e} \arctg \frac{2m}{k_e} \cdot D^{(l)}(k_e^2) \right] k_e^2 dk_e. \qquad (7)$$

For further calculations it is necessary to concretize the form of the function $D^{(l)}(k_e^2)$. It is impossible to take it in the usual form $D^{(l)}(k_e^2) = \frac{1}{ik_e^2}$ since this leads to the divergence of the integral at its lower bound ($k_e = 0$). Taking the expression for the gauge function $D^{(l)}(k_e^2)$ which arises at $k_e \to 0$ more slowly than $1/k_e^2$ we shall get for the electromagnetic mass

$$\delta m = \frac{\alpha}{(2\pi)^2 N} \int_0^{u_{max}} M(u) du, \qquad (8)$$

where

$$M(u) = \left(1 + \frac{4-u^2}{2u} \operatorname{arctg} \frac{2}{u}\right) \frac{u}{u + \alpha/(8m)} + M_1(u), \qquad (8a)$$

$$M_1(u) = m^3 u D^{(l)}(m^2 u^2), \quad u_{max} = k_{max}/m = \lambda_c/r_0.$$

Further we shall have interest in the behaviour of the kernel $M(u)$ at $u \gg 1$ and $u \ll 1$. It is possible to choose the function $D^{(l)}$ in such a manner that it will not influence on the behaviour of the kernel $M(u)$ in the denoted domains of the variable $u$. The gauges in which the functions $D^{(l)}$ have denoted properties we shall call further by the generalized Landau gauges (examples of such gauges are given in [10,11]).

4.1) Let the Compton wavelength surpasses the model characteristic parameters ,i.e. $\lambda_c \gg r_0$, $\lambda_c \gg 8/\alpha$ and from the sense of those scales it follows that $r_0 \ll 8/\alpha$. If one acts in the framework of generalized Landau gauges the main contribution to the integral (7) is given by the domain $1 < u \ll \alpha/8m$. In this case we have

$$M(u) \approx \frac{16}{3} \frac{1}{u(u + \alpha/(8m))}, \qquad (8b)$$

and thus

$$\delta m = \frac{\alpha}{(2\pi^2)N} \int_1^{\alpha/8m} M(u) du = \frac{32}{3\pi^2 N} m \ln \frac{\alpha}{8m}. \qquad (9)$$

It is the quantum result for the electromagnetic fermion mass at the first nonvanishing order of this quantity expansion into the parameter $N^{-1}$ (we remark that in distinction from $QED_4$ here is obtained the convergence result for the electromagnetic mass)..

2) Now we shall consider the case $\lambda_c \ll r_0$, $\lambda_c \ll 8/\alpha$. In this case the integration in (8) is performed upon the domain $0 \ll u \ll (mr_0)^{-1}$. Using generalized Landau gauge we obtain

$$M(u) \approx \frac{\pi}{u + \alpha/(8m)}, \qquad (10)$$

$$\delta m = \frac{\alpha}{4\pi N} \ln \frac{8}{\alpha r_0}. \qquad (10a)$$

If $\delta m = m$ this result was obtained from the classical expression for the fermion self mass [4,5]. In the quantum case, if the fermion mass is completely electromagnetic we have

$$m = \frac{\alpha}{8} e^{-\frac{3\pi^2 N}{32}}. \qquad (9a)$$

The conditions $m \ll \alpha/8$ in the quantum case and $m \gg \alpha/8$ in the classical one are fulfilled with accuracy which is high enough. In the quantum case when $\delta m = m$, it is necessary to take for $N$ the value $N_c \approx 3 \div 4$ [8,12,13] and the mass has an exponential character of smallness upon $N$. With accuracy to the numeral multiplier at the exponent index at Landau gauge this result was obtained in [9]. A small difference between our result and that given in [9] is connected with that in this work are preserved not all terms of the same order upon $u$, giving the main contribution to the integral in the domain $1 \ll u \ll \frac{\alpha}{8m}$, if the kernel (8) expansion into power series upon $u$ is carried out. As for the classical result, the condition $m \gg \alpha/8$ fulfils if the unequality $\ln(8/(\alpha r_0)) \gg 1$ satisfies, that is the more strong restriction than $(8/(\alpha r_0)) \gg 1$.

5. Thus, if one proceeds from the expression for the fermion self energy (1) in QED$_3$ at the first nonvanishing order of $N^{-1}$- approximation, it is possible to obtain the quantum result (9) as well the classical one (10a). The condition $\lambda_c \gg a$ corresponds to the quantum case whilst the condition $\lambda_c \ll a$ corresponds to the classical one ($a$ - denotes the characteristic parameters of the model; in QED$_3$ they are $r_0$ and $8/\alpha$). It is impossible to consider the last case as a completely classical one because the classical expression for the electromagnetic energy on the plane

$$\delta m = \int T_{44}(\vec{r}) d^2\vec{r} = \frac{1}{2} \int E^2 d^2\vec{r}, \qquad (11)$$

uses the source strength value involving the vacuum polarization effects, as it was noticed in p.2.

The following features of the obtained results should be emphasized:

1) The classical expression for the particle electromagnetic mass in our case has higher degree of singularity if $r_0 \to 0$ than the quantum one. Such a situation takes place at the first nonvanishing order of the perturbation theory in QED$_4$.

2) The fermion electromagnetic mass evaluated in [4,5] from the classical expression (11) coincides with (10a) and does not depend on gauge. Meanwhile, if one uses the self energy integral (1) as it follows from the discussion given at p.3 of our work the expression obtained by using of this integral coincides with the classical one at definite gauge only. In other words in the quantum case the fermion electromagnetic mass depends on gauge that, from our point of view, is not a serious problem since this quantity is not observed in the experiment. The more essential remark concerns the full mass. The dependence of this quantity on gauge in our case by the quantum approach is possible to explain by the shortage of approximation. Really the dynamical mass of the fermion arises in consequence of the chiral symmetry breaking of the initial Lagrangian of the model. As it turns out by investigating the simplest nonperturbative approximations in $QED_3$ and $QED_4$ [14-16] the denoted process takes place only under the condition when the models dimensionless parameters (nonrenormalized coupling constant $\alpha_0$ in $QED_4$ and fermions number $N$ in $QED_3$) in the metioned approximations depend on gauge. Along with it as it follows from the works [12,13,17,18] the improvement of the approximation in some measure diminishes the influence of the gauge choice at the critical values of these parameters.

    We shall notice also that the improvement of the approximation is necessary for the proving that the coinciding of the classical results obtained with the aid of the relations (1) and (11) is not an approximation artefact, and it remains beyond the bounds of the first nonvanishing order of the $N^{-1}$- expansion.

    We have to notice that for more complete studying of the problem, it is necessary to proceed from the nonlinear Lagrangian in $QED_3$, in this case the form of the component $T_{44}$ changes itself in comparison with the linear one. As it is noticed in [19], in $QED_4$ the using of the nonlinear Lagrangian changes essentially the asymptotic of the point source field potential by $r \to \infty$ compared with the obtained by the evaluation of this potential taking into account the first nonvanishing term of the $N^{-1}$- expansion of the polarization operator. There are reasons to expect that the same situation takes place in $QED_3$.

## References

1. Schweber S. An Introduction to Relativistic Quantum Field Theory. Row,

  . Peterson and Co∗Evanston, ILL., Elmsford, N.Y. 1961.

2. Fomin P.I. //Physics of the Elementary Particles and Atomic Nucleus, 1976, №7, 687.

3. Efimov G. V. //Nonlocal Interaction of the Quantum Field. Moscow.Nauka, 1977, 367 p.

4. Pevzner M.Sh.. // Rus. Phys. Journ. 1998,№11,112.
5. Pevzner M.Sh. //Ukr. Phys. Journ. 1999, **44**,1294.
6  Burden C.J., Praschifka J. and Roberts C.D. //Phys. Rev. 1992, **D46,** 2695.
7.   Maris P. //hep–ph/9508323.
8.   Gusynin V.P., Hams A.H. and Reenders M. // Phys. Rev. 1996**, D53**, 2227.
9.   Pisarski R. //Phys. Rev. 1984, **D29,**. 2423..
10.   Simmon E.H. //Phys. Rev. 1990, **D42,** 2933, Kugo T.and Mitchard. M.G.// Phys. Lett.1992, **B282,** 162.
11. Pevzner M.Sh.. // Rus. Phys. Journ. 1999, №9,91.
12.   Daniel Nash //Phys. Rev. Lett.1989, **62,**3024..
13.   Maris P. //hep–ph 9606214.
14.   Fomin P.I., Gusynin V.P., Miransky V.A. and Sitenko Yu.A. //Riv Nuovo Cim. 1983, **6**, 1.
15.   Miransky V.A. //Nuovo Cim.1985,. **90A,** 148.
16.   Appelquist Thomas. //Suppl. Progr. Theor. Phys.1985, **85,** 244..
17.   Atkinson D. , Bloch J.C.R., Gusynin V. P., Pennington M. R., Reenders M.R. Strong QED with Weak Gauge Dependence: Chiral Coupling and Anomalous Dimensions.  RUG–TH–930901; DTP–931/62.
18.    Bashir A. and  Pennington M.R. Gauge Independent Chiral Symmetry Breaking.–Preprint.–DTP–94/98.
19.   Achiezer A. I. and Berestetsky V.B. //Quantum Electrodynamics. Moscow.Nauka, 1981, 432 p.